\documentstyle[12pt,epsf,epsfig]{article}

\def\ga{\mathrel{\raise.3ex\hbox{$>$\kern-.75em\lower1ex\hbox{$\sim$}}}}

\def\la{\mathrel{\raise.3ex\hbox{$<$\kern-.75em\lower1ex\hbox{$\sim$}}}}

\begin{document}

\begin{center}

{\large\bf The Genesis of Cosmological Tracker Fields}

\vskip0.4in

Vinod B.Johri$\dagger$

\vskip0.2in

{\it
Theoretical Physics Institute,School of Physics and Astronomy,\\
University of Minnesota, Minneapolis, MN 55455,USA}

\vskip0.2in

{\bf Abstract }

\end{center}

{\small The role of the quintessence field as a probable candidate
for the repulsive dark energy, the conditions for tracking and the
requisites for tracker fields are examined. The concept of `integrated
tracking' is introduced and a new criterion for the existence of tracker
potentials is derived assuming monotonic increase in the scalar energy 
density parameter $\Omega_\phi$ with the evolution of the universe 
as suggested by the astrophysical constraints. It provides a
technique to investigate generic
potentials of the tracker fields. The general properties of the tracker 
fields are discussed and their behaviour with respect to tracking
parameter $\epsilon$ is analyzed.
It is shown that the tracker fields around the
limiting value $\epsilon \simeq \frac 23$ give the best fit with
the observational constraints.}

PACS numbers: 98.80.Cq, 98.65.Dx, 98.70.Vc

\vskip 0.2 in

There is strong evidence, based on recent luminosity-redshift 
observations of Type $I_a$ supernovae [1] and consistently low
measurements of matter density [2], to suggest that the major
fraction of the energy content of the observable universe 
consists of an 'exotic matter' with negative pressure, often
referred to as 'dark energy'[3]. In cold dark matter(CDM) cosmology,
the most probable candidates for dark energy are the
cosmological
constant $\Lambda$ and the weakly coupled scalar fields 
with negative pressure which might mimic $\Lambda$ to produce
enough repulsive force to counter gravitational attraction and
cause acceleration in the expanding universe at the present epoch.
Comprehensive review reports on the observational,theoretical,physical and
anthropic significance of the cosmological constant $\Lambda$ have
been published by Zeldovich [4], Weinberg [5], Sahni and Starobinsky [6].
The cosmological constant seems to be a natural choice for the 
source of cosmic repulsion but it is hard to reconcile its constant
value $\Lambda \sim 10^{-47} {GeV^4}$ (to be comparable with the
present energy density of the universe) with the particle physics
scales $ \sim 10^{56}{GeV}^4$ subsequent to inflation. Since
$\Lambda$ has stayed constant through cosmic evolution, it demands
setting up a new energy scale to explain as to why it should
take 15 billion years of time for $\Lambda$ to dominate in the
universe today (known as coincidence problem). To address this problem,
a comprehensive study of the observational consequences of a dynamical
$\Lambda$ term (representing vacuum energy) decaying with time [7] 
and the cosmological consequences of rolling 
scalar fields was undertaken [8]; subsequently, Caldwell et al [12]
discussed the possibility that a significant contribution to the
energy density of the universe might be from the scalar fields
with an evolving equation of state, unlike radiation, matter or 
$\Lambda$ fields and proposed the nomenclature 'quintessence'
for such scalar fields which, during the
process of roll-down, acquire negative pressure and might act as
$\Lambda_{eff}$. But for the scalar energy density
$\rho_{\phi}$ or $\Lambda_{eff}$ to be comparable with the present
energy density $\rho_n$ of the universe, the initial conditions
for the quintessence fields must be set up carefully and fine
tuned. To overcome the 'fine tuning' or the 'initial value'
problem, the notion of tracker fields [13,14] was introduced. It
permits the quintessence fields with a wide range of
initial values of $\rho_{\phi}$ to
roll down along a common evolutionary track with $\rho_n$ and 
end up in the observable universe with $\rho_{\phi}$ comparable
to $\rho_n$ at the present epoch. Thus, the tracker fields can get
around both the coincidence
problem and the fine tuning problem without the need for defining
a new energy scale for $\Lambda_{eff}$.
Although tracking is a useful tool to promote quintessence as a likely
source of the Ômissing energyÕ in the universe, the concept of tracking 
as given by Steinhardt et al [13,14] does not ensure the physical viability 
of quintessence in the observable universe. It simply provides for synchronized
scaling of the scalar field with the matter/radiation field in the expanding
universe in such a way that at some stage (undefined and unrelated to
observations), the scalar field energy starts dominating over matter and may
induce acceleration in the hubble expansion. Since there is no control over
the slow roll-down and the growth of the scalar field energy during tracking,
the transition to the scalar field dominated phase may take place much later than observed. Moreover, any additional contribution to the energy density
of
the universe, such as quintessence, is bound to affect the dynamics of 
expansion and structure formation in the universe. As such, any physically viable scalar field must comply with the cosmological observations related to
helium abundance, cosmic microwave background and galaxy formation, which are the pillars of the success of the standard cosmological model. A
realistic theory of tracking of scalar fields must, therefore, take into account the 
astrophysical constraints arising from the cosmological observations. With this perspective in mind, we have introduced the notion of `integrated
tracking' in
this paper which implies tracking compatible with astrophysical constraints. It
provides a firm and credible foundation to the quintessence theory. Most of the investigations [10,11,12,13] on the scalar fields so far have been confined to
exploring scalar
potentials which roll down with the desired tracking behaviour to end up with dominance of quintessence energy. The first theoretical derivation of 
the tracking condition was attempted by Steinhardt et al [14] who put
forth different criteria for tracking under varying conditions and
discussed the tracking properties of certain exponential and inverse
power law potentials. In this Letter, we report our investigations
towards a systematic theory of Ôintegrated trackingÕ in which the tracking 
behaviour of the scalar field is closely related to the growth of its
cosmological density parameter through the tracking parameter $\epsilon$,
subject to astrophysical constraints as discussed in the paper.
This approach provides us a powerful technique to study the general
behaviour of the tracker fields with respect to $\epsilon$ and also
a window to investigate the generic potentials of the tracker fields
instead of dealing with isolated potentials and their properties.

In general, the energy densities $\rho_n$ and $\rho_{\phi}$
scale down at different rates in the expanding universe. For a
scalar field with potential $V(\phi) = \frac{1}{2}(\rho_{\phi} - 
p_{\phi})$ and kinetic energy $\frac{1}{2}{\dot\phi^2}$ =
$\frac{1}{2}(\rho_{\phi}+p_{\phi})$,the equation of motion of
the scalar field
\begin{equation}
\ddot\phi+3H\dot\phi+V^{\prime}(\phi) = 0
\end{equation}
leads to $\rho_{\phi} \sim a^{-3(1+w_\phi)}$
where $1+w_{\phi} =
1+\frac{p_{\phi}}{\rho_{\phi}}=\frac{\dot\phi^2}{\rho_{\phi}} 
= 2\eta$, $\eta$ denotes the ratio of kinetic energy to $\rho_\phi$.

For the background energy,
$\rho_n\sim a^{-3(1+w_n)}\sim\frac{1}{a^n}$, $n=4$(radiation)
and $n=3$(matter). 
Obviously, the scalar field has a wider range of scaling as
$\rho_{\phi} \sim a^{-6\eta}$, $(0\leq\eta\leq 1)$ depending 
upon the choice of $\eta$. When the
kinetic energy is dominant, $\rho_{\phi}$ can scale down as steeply
as $\frac{1}{a^6}$. It rolls down slowly as $V(\phi)$ starts
dominating and the rolling reduces to a crawl as $\eta$ tends to
zero. Therefore, the kinetic energy plays an important role in
scaling down the energy of the scalar field. In order to solve
the 'dark energy' problem, we want domination of the scalar field
($\rho_\phi\ga \rho_n$) today but at the same time it is
imperative that $\rho_\phi < \rho_n$ during radiation and
matter dominated era and grows slowly to the present state so as
not to interfere with the formation of galactic structure and
the success of nucleosynthesis during cosmic evolution.
Therefore, tracking requires proper synchronization of the 
scaling of the two fields so that $\rho_\phi$ rolls down slower
than $\rho_n$ (i.e $w_\phi< w_n$) along a common evolutionary
track and eventually overtakes it, causing acceleration in the
cosmic expansion. This implies that $w_\phi<\frac13$ during
radiation era, $w_\phi<0$ during matter domination and $w_\phi$
tends to $-1$ during scalar energy dominated phase, constraining
$\eta$ to lie in the range $(0\leq\eta<\frac23)$.
Naturally a fixed value of $\eta$ does not lead to tracking.
It must vary with the roll down of the scalar field but its
variation over the wide span of the cosmic time is so small 
that it may be regarded as almost a constant and the time
derivatives of $\eta$ may be neglected. This assumption
simplifies the dynamics of evolution of the tracker fields.
Using Eq. (1), the logarithmic differentiation of $V(\phi) =
(1-\eta)\rho_\phi$ yields an important condition to be satisfied
by the tracker fields. 
\begin{equation}
\pm\frac{V'(\phi)}{V(\phi)} = 6\eta\frac{H}{\dot\phi}
=\frac{\sqrt{6\eta}}{M_p\sqrt{\Omega_\phi}} 
\end{equation}
where $\Omega_\phi \equiv \frac{\rho_{\phi}}{\rho_\phi+\rho_n}$, 
$H$ is the Hubble constant given by
$H^2=\frac{\rho_\phi+\rho_n}{3M_p^2}$ and $M_p=2.4\times10^{18} 
GeV$ is
the reduced Planck mass. According to our notation, the prime
denotes derivative with respect to $\phi$, an overdot denotes
time-derivative and $\pm$ sign applies to $V'>0$ and $V'<0$
respectively.

It is remarkable that the maintenance of the tracker condition (2)
constrains the
order of magnitude of the various terms involved. For instance,
the middle term $\sqrt{\frac{\rho_\phi+\rho_n}{\rho_\phi}}$ in
(2) must remain nearly of $O(1)$ throughout tracking. It
follows, therefore, that the scalar fields with
$\rho_\phi\ll\rho_n$ will remain frozen until hubble expansion
slows down to the level when $\rho_n\approx\rho_\phi$; thereafter,
the tracker condition holds good and tracking takes place. Thus,
the tracker condition ensures that a large class of quintessence
fields with widely diverse initial conditions
$(\rho_\phi\ll\rho_n)$ would scale down to the same present state
with $\rho_\phi\simeq\rho_n$. However, the scalar fields with
$\rho_\phi\gg\rho_n$ violate astrophysical constraints given below although
relation (2) continues to hold good. Hence the tracker condition
(2) is not a sufficient condition for tracking and it has to be
supplemented with more stringent requirements based on the
astrophysical constraints discussed below.

Observationally, the cosmological density parameter $\Omega_\phi$ 
of the scalar field is an important quantity since
it measures the relative magnitude of the energy
densities $\frac{\rho_\phi}{\rho_n}$ during cosmic evolution
as given by
\begin{equation}
\Omega_\phi = (1+a^{-3\epsilon})^{-1}
\end{equation}
where $\epsilon\equiv w_n-w_\phi$\,, $0<\epsilon\leq1$.
It may be used to regulate the tracking behaviour according to the
following astrophysical constraints:\\[4mm]
I. $\Omega_\phi < 0.13 - 0.2$ at the nucleosynthesis epoch [10]
around redshift $z=10^{10}$ \\[1mm]
II. $\Omega_\phi < 0.5$ during galaxy formation epoch [3] 
around $z=2$ to 4 \\[1mm]
III. $\Omega_\phi = 0.5$ at the onset of acceleration in
cosmic expansion at redshift $z\,(0\leq z <2)$ \\[1mm] 
IV. $\Omega_\phi \simeq 0.65\pm\,0.05$ with $w_\phi\leq-0.4$ at 
the present epoch ($z=0$) [15]\\[4mm]
\indent 
As stated above, the tracker condition given by Eq.(2) ensures slow rolling of the scalar potential $V(\phi)$ and may be regarded as a necessary condition
for tracking but not as a criterion for tracker fields. To lay down a physical 
criterion for tracker fields, we take a clue from the astrophysical constraints 
I - IV which require progressive growth of $\Omega_\phi$ during tracking and postulate that $\dot\Omega_\phi > 0$ for tracker fields. This can be
monitored by a single parameter $\epsilon$ (known as tracking parameter) since it reveals a clear picture of scaling of $\rho_\phi$
vs. $\rho_n$ throughout the range of tracking. The limiting value of $\epsilon$, as derived from Eq.(8) ensures the transition from matter to the scalar
dominated phase.

By logarithmic differentiation of Eq. (2),
$\frac{\dot\Omega_\phi}{\Omega_\phi}$ may be expressed in
terms of $V(\phi)$ and its derivatives as
\begin{equation}
\frac{\dot\Omega_\phi}{\Omega_\phi} = \mp\,12\eta H(\Gamma-1)
\end{equation}
where $\Gamma \equiv\frac{V''V}{V'^2}$.
Again the time derivative of Eq.(3) gives
\begin{equation}
\dot\epsilon\,\ ln\,a =
\frac13\,\frac{\dot\Omega_\phi}{\Omega_\phi\Omega_n} -
\epsilon H\,.
\end{equation}
The choice of $V(\phi)$ may
be further restricted by setting a stronger condition for tracking i.e
$\dot\epsilon\geq 0$ which requires
\begin{equation}
\dot\Omega_\phi \,\geq 3\epsilon H\Omega_\phi\Omega_n > 0\,.
\end{equation}
Eqs. (4) and (6) lead to the final criterion for tracker fields
\begin{equation}
\mp(\Gamma - 1) \,\geq \frac{\epsilon\Omega_n}{4\eta}\,.
\end{equation}
{\it The above inequality implies that a given scalar
potential $V(\phi)$ will give rise to a tracker field if
$\Gamma\leq 1-\frac{\epsilon\Omega_n}{4\eta}$ in case of 
increasing potential ($VÕ>0$) and
$\Gamma \geq 1+ \frac{\epsilon\Omega_n}{4\eta}$ in case of 
decreasing potential ($VÕ<0$)
where $\frac{\epsilon\Omega_n}{4\eta} <1$ and $\epsilon$ conforms
to the astrophysical constraints I - IV.}

The tracking criterion for the quintessence potentials derived 
above is based on the restrictive assumption that $\Omega_\phi$ 
increases monotonically through most of the cosmological history. 
In fact, this assumption is motivated by the astrophysical 
Constraints I - IV listed above which demand the progressive 
increase in the fractional magnitude of $\Omega_\phi$ from 
nucleosynthesis epoch around $z\simeq 10^{10}$ through galaxy 
formation era to the present day (z=0). This assumption, although
restrictive, seems to be 
quite natural and consistent with the thermal
history of the universe. Dodelson et al [17] have suggested an 
alternative scenario for Ôtracking with oscillating energyÕ under 
which the scalar potential with sinusoidal modulation oscillates 
about the ambient energy density. The oscillating tracker potentials 
may satisfy the astrophysical constraints I - IV provided the magnitude 
and frequency of the oscillations are fine-tuned to comply with the 
specific requirements listed under the constraints.
Similar existence conditions for the tracker fields have been
obtained by Steinhardt et al [14] but they hold under the
restriction $\Omega_n = 1$. The above results
not only lay down the criterion for the existence of
tracker fields but they also emphasize the importance of the
tracker parameter $\epsilon$ which may be used with advantage to
derive generic potentials for tracking fields as discussed
below. Again, the astrophysical constraints during the phase
transition from matter to scalar dominated phase, when
$\Omega_n$ is significantly less than 1, may be utilized to
limit the range of $\epsilon$ as shown below. This is to ensure
that the roll down of the scalar field is not too slow and the
universe must enter the phase of accelerating expansion at the
right epoch at red shift $z_0\, (0<z_0<2)$ after the galactic
structure has formed.

The general behaviour of tracker fields, regardless of the
form of $V(\phi)$, may be outlined by deducing the value 
of $\epsilon$ at the onset of acceleration subject to
astrophysical constraints III and IV. Using the Friedmann 
equation during matter dominated era
\begin{equation}
\frac{2\ddot a}{a}\, =
-\frac{\rho_\phi[1+3w_\phi+(\rho_n/\rho_\phi)]}{3M_p^2},
\end{equation}
the condition for onset of acceleration i.e. $\ddot a\ga 0$ when
$\frac{\rho_n}{\rho_\phi}\simeq 1$, leads to the limiting value
$\epsilon_0\ga\frac23$ around the transition to the scalar
dominated phase. The relation $\frac{\rho_\phi}{\rho_n} =
2(1+z)^{-3\epsilon}$ involving redshift $z$, enables us to find
$\rho_\phi$ at various landmark epochs in cosmic evolution for
different values of $\epsilon$ and examine which of these lead
to desired tracking behaviour.
\begin{figure}
\begin{center}
\mbox{\epsfig{file=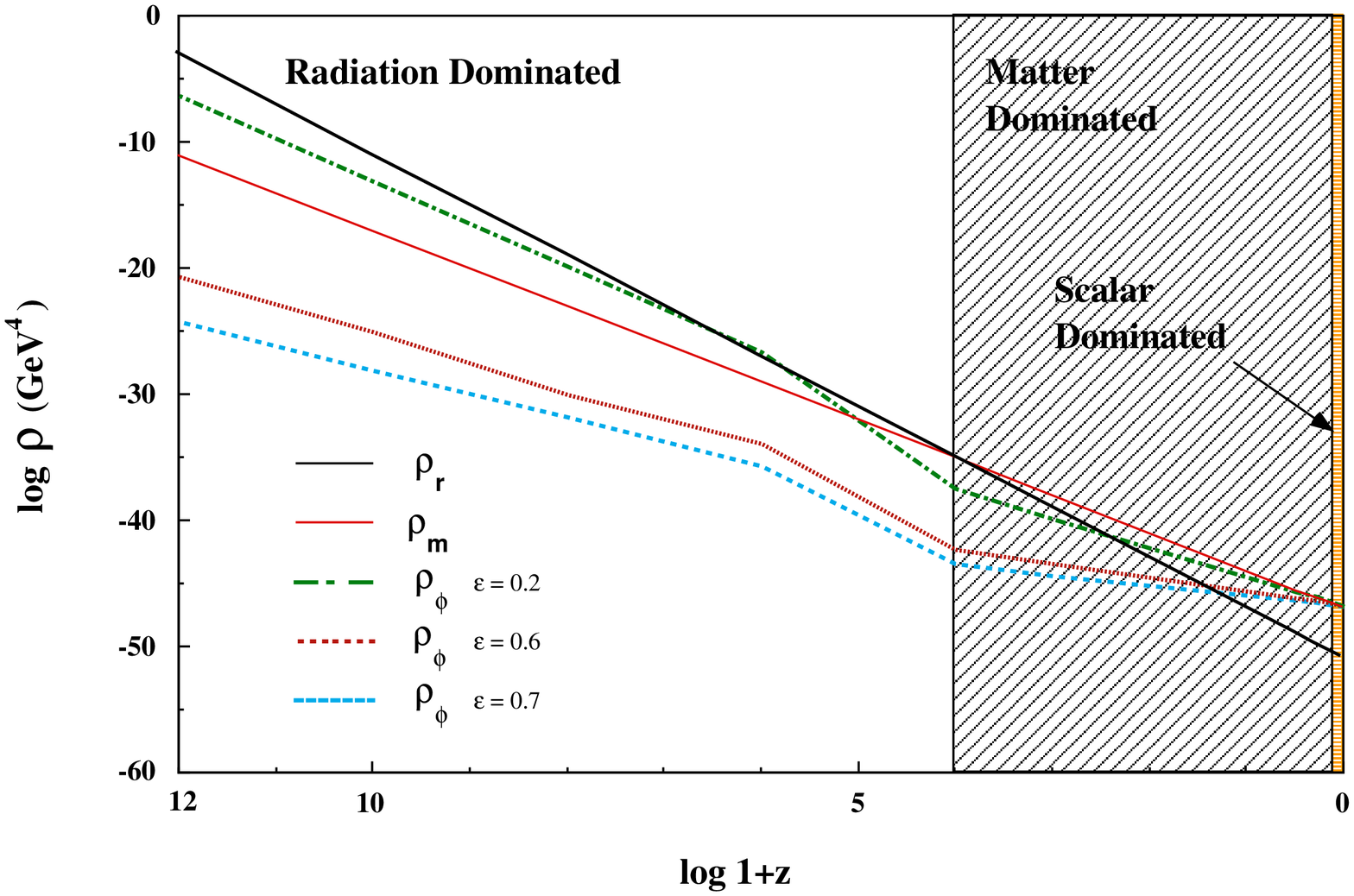,height=9.5cm,width=14cm}}
\end{center}
\caption[.]{\label{fig:bsg1}\it
Scaling of energy densities $\rho_r$ (radiation), $\rho_m$ (matter)
and $\rho_\phi$ (tracker field) $vs.$ red shift $z$ in the expanding
universe. The curves are plotted by taking values of $\epsilon$ at the 
point of transition to the scalar dominated phase.}
\end{figure}
By choosing $\epsilon$ = 0.2, 0.6 and 0.7 successively in the 
redshift relation, we have shown
by the indexed curves in the figure as to how the relative scaling
of $\rho_r$, $\rho_m$ and $\rho_\phi$ takes place
during different phases of evolution in the expanding universe.
In fact, $0<\epsilon<\epsilon_0$ during matter and radiation
dominated era; as such $\rho_\phi$ would track down closer to
$\rho_n$ than shown in the figure. It is found that the models
with $\epsilon = 0.6$ and $0.7$ satisfy all the requisite
astrophysical constraints whereas $\epsilon = 0.2$ violates II
and III. Therefore,the
best-fit quintessence models correspond to $\epsilon\approx 0.66$
as also indicated by the limiting value derived above. This is
also consistent with the concordance analysis [15], based on a
comprehensive study of observational constraints on spatially
flat cosmological models containing a mixture of matter and
quintessence.

For the known functional values of $\epsilon(\phi)$, the generic 
potentials for the tracker fields can be found by putting the tracker 
criterion in the form
\begin{equation}
\pm\frac{V''V - V'^2}{V'^2}\,\geq
\frac{\epsilon(1-\Omega_\phi)}{2(1+w_n-\epsilon)}
\end{equation}
Inserting the value of $\Omega_\phi$ from Eq.(2), we get on simplification
\begin{equation}
\pm\frac{\zeta'}{\zeta^2 - k^2}\geq \,f(\phi)\equiv 
\frac{\epsilon}{2(1+w_n-\epsilon)}
\end{equation}
where $\zeta\equiv V'/V$ and $k^2=\frac{6\eta}{M_p^2}$.
The above equation yields generic potential $V(\phi)$ for suitable choice of
$f(\phi)$.
For example, $\epsilon = 0$ corresponds to the generic potentials of the form 
$V(\phi) \sim exp[\beta\phi]$ for which $\Omega_\phi$ remains
constant throughout. Astrophysical constraints I and II may be
satisfied by choosing $\Omega_\phi <$ 0.15 as discussed in
[14] but the tracking remains incomplete since there is
no onset of acceleration as required by the constraint III. If
$\epsilon$ = constant throughout, there is
limited tracking since the ratio of the kinetic energy to the
potential energy of the scalar field remains stationary
throughout rolling with a ceiling fixed by the
constraint II; further $V(\phi)$ never attains a constant
value to play the role of $\Lambda$ in the universe. The
generic potentials are of the hyperbolic form; in the particular
case of matter dominated universe $(\Omega_n\simeq1)$, the potentials
are of the inverse power law form.
Power law potentials have been discussed extensively by several authors
[8,10,11,13,14,16].

The detailed investigations involving mathematical theory of tracker fields,
the stability and astrophysical consequences of tracker solutions will be
published under a separate communication elsewhere.

The author gratefully acknowledges useful discussions with
Keith Olive and Panagiota Kanti and their keen interest and help in the
 preparation of this manuscript.

$\dagger$Permanent address: Department of Mathematics and
Astronomy, Lucknow University, Lucknow 226007, India. Email:
vinodjohri@hotmail.com

\end{document}